# A Quantum Like Interpretation and Solution of Einstein, Podolsky, and Rosen Paradox in Quantum Mechanics

## Elio Conte

*Department of Pharmacology and Human Physiology and Tires, Center for Innovative Technologies for Signal Detection and Processing, University of Bari, Italy;*
*School of Advanced International Studies on Theoretical and nonLinear Methodologies, Bari, Italy*

**Abstract:**
We discuss the Einstein, Podolsky, and Rosen paradox as it was formulated by Asher Peres in 1992 [4]. On this basis we realize an algebraic quantum like elaboration showing that in this formulation the paradox may be still interpreted and solved.

## 1. Introduction

Definition of Quantum Like may assume different interpretations according to various authors. In the present paper we will intend by this term a rough scheme of quantum mechanics, a bare bone skeleton of the fundamental theory, as it may be introduced by using the Clifford algebra. It was discussed in a number of previous papers [1] so that we will no more consider it in the present work and the reader is send back to such papers in order to deepen this argument. Also our exposition and interpretation of the EPR paradox was previously discussed elsewhere in a preliminary form [2].
Therefore we will state here only some quantum like notations just to enable the reader to follow our subject.

## 2. A rough Quantum Like Scheme of Quantum Mechanics

Let us introduce two basic algebraic axioms:

1) $e_i$, $i = 1,2,3$ are abstract – symbolic-algebraic elements whose square gives 1,

$$e_i^2 = 1.  \qquad (1)$$

2) Such algebraic elements are anti commutative, that is to say :

$$e_i e_j = -e_j e_i, \qquad i = 1,2,3; j = 1,2,3; i \neq j. \qquad (2)$$

Only these two algebraic axioms, (1) and (2), are required [1, 2] in order to realize a rough quantum mechanical scheme.
Let us comment on the axioms (1) and (2). Here the problem is only of interpretation. We call (1) the axiom of the potentiality. In fact, owing to the axioms (1) and (2), the $e_i$ cannot represent any number in some field. However, owing to the axiom (1), each $e_i$ in (1) has the potentiality that we could attribute to it a numerical value, that is or +1 or -1.
Usual quantum mechanics also introduces a net distinction between potentiality and actualization from its starting and, in particular, through the superposition principle of quantum states.
A theorem discussed in [1] demonstrates that the algebra given by (1) and (2) actually exists and this is to say that the two given axioms are sufficient to realize an algebraic structure that we call the Algebra A. According to our interpretation in some manner it represents the algebra of the potentiality. The sense of this notion is that we never may give a direct numerical value to the algebraic elements or members of this algebra but, however, such mathematical structure preserves at any time the potentiality that some numerical value could be attributed to its basic elements. The theorem 2, discussed in [1], demonstrates that we may also realize the passage from potentiality to actualization in the sense that we are also in the condition to attribute some definite numerical values to the basic algebraic elements identified in (1) and (2), and in this case we have a passage from the Algebra A to a subalgebra B where now to one and only to one element $e_i$, we may attribute a definite numerical value, say o +1 or -1. In this manner, by theorem 2, we describe the passage from potentiality to actualization.

In quantum mechanics we have the passage from potentiality to actualization through the so called mechanism of wave function reduction that, however, it is only admitted but never derived in the framework of such theory.

Let us see some other features in such algebraic scheme. In algebra A, considering that each element $e_i$ has the potentiality to assume the numerical value of +1 or of -1, we are implicitly admitting that such $e_i$ are characterized from an intrinsic indetermination. This is to say that at the stage of their potentiality there exists an ontological probability $p_{+1}$ for each $e_i$ to assume the value +1 and a probability $p_{-1}$ to assume the value -1, given in the following manner:

$$p_{+1} = \frac{1}{2} + <e_i> \quad \text{and} \quad p_{-1} = \frac{1}{2} - <e_i> \qquad (3)$$

being $<e_i>$ the mean value for the basic element $e_i$.

We see that in this manner we have delineated a rough quantum scheme through an algebraic structure. Starting the general validity of quantum mechanics, the axioms (1) and (2) could represent the mathematical counterpart of some basic features of our reality, that one of potentiality. Let us add still some other mathematical notation.

The isomorphism of such basic elements with matrices at $n = 2$ is well known. In the case of the (1) and the (2) they are given in the following manner:

$$e_1 = \begin{pmatrix} 0 & 1 \\ 1 & 0 \end{pmatrix}, \quad e_2 = \begin{pmatrix} 0 & -i \\ i & 0 \end{pmatrix} \quad \text{ed} \quad e_3 = \begin{pmatrix} 1 & 0 \\ 0 & -1 \end{pmatrix} \qquad (4)$$

By using direct product of matrices we may realize sets of basic elements at different orders n. As example, at n = 4, we have that

$$E_{01} = 1 \otimes e_1 = \begin{pmatrix} 1 & 0 \\ 0 & 1 \end{pmatrix} \otimes \begin{pmatrix} 0 & 1 \\ 1 & 0 \end{pmatrix} = \begin{pmatrix} 0 & 1 & 0 & 0 \\ 1 & 0 & 0 & 0 \\ 0 & 0 & 0 & 1 \\ 0 & 0 & 1 & 0 \end{pmatrix}, \quad E_{10} = e_1 \otimes 1 = \begin{pmatrix} 0 & 1 \\ 1 & 0 \end{pmatrix} \otimes \begin{pmatrix} 1 & 0 \\ 0 & 1 \end{pmatrix} = \begin{pmatrix} 0 & 0 & 1 & 0 \\ 0 & 0 & 0 & 1 \\ 1 & 0 & 0 & 0 \\ 0 & 1 & 0 & 0 \end{pmatrix}$$

$$E_{02} = \begin{pmatrix} 0 & -i & 0 & 0 \\ i & 0 & 0 & 0 \\ 0 & 0 & 0 & -i \\ 0 & 0 & i & 0 \end{pmatrix}, \qquad E_{03} = \begin{pmatrix} 1 & 0 & 0 & 0 \\ 0 & -1 & 0 & 0 \\ 0 & 0 & 1 & 0 \\ 0 & 0 & 0 & -1 \end{pmatrix},$$

$$E_{20} = \begin{pmatrix} 0 & 0 & -i & 0 \\ 0 & 0 & 0 & -i \\ i & 0 & 0 & 0 \\ 0 & i & 0 & 0 \end{pmatrix}, \qquad E_{30} = \begin{pmatrix} 1 & 0 & 0 & 0 \\ 0 & 1 & 0 & 0 \\ 0 & 0 & -1 & 0 \\ 0 & 0 & 0 & -1 \end{pmatrix} \qquad (5)$$

In correspondence of the (1) and the (2) we have now that
$E_{01}^2 = E_{02}^2 = E_{03}^2 = 1$ and $E_{01}E_{02} = iE_{03}$, $E_{02}E_{03} = iE_{01}$, $E_{03}E_{01} = iE_{02}$
and equivalent relations for $E_{j0}$, $j = 1,2,3$. (6)

Generally speaking, we may consider the following possible basic sets at order n=4 :
$(E_{01},E_{12},E_{13}),(E_{01},E_{22},E_{23}),(E_{01},E_{32},E_{33}),(E_{02},E_{11},E_{13}),(E_{02},E_{21},E_{23}),(E_{02},E_{31},E_{33}),(E_{03},E_{11},E_{12}),$

$(E_{03},E_{21},E_{22}),(E_{03},E_{31},E_{32}),(E_{10},E_{23},E_{33}),(E_{10},E_{22},E_{32}),(E_{10},E_{21},E_{31}),(E_{20},E_{12},E_{32}),(E_{20},E_{11},E_{31}),$

$(E_{30},E_{13},E_{23}),(E_{30},E_{12},E_{22}),(E_{30},E_{11},E_{21})$. (7)

These are the possible basic sets at this order (n=4) with $E_{ij} = E_{i0}E_{0j}$. It is seen that each basic set is regulated by the basic permutation $(i\ j\ k)$ of $(1,2,3)$. Each set identifies an abstract space that would be the space of quantum like events in which potentiality, expressed by the (1) and the (2), or equivalently, by the (6), operates. We should have the algebraic PRESPACE of potentialities actualized after in the ordinary space of our experience.

### 3. The Paradox of Einstein, Podolsky and Rosen.

As it is known, there are several versions of the paradox, starting with the initial EPR formulation of the authors in 1935 [3].
In this paper we will follow the excellent formulation that was given by Asher Peres in 1992 [4].
We quote directly from this article:
*A fundamental issue was raised by Einstein, Podolsky, and Rosen in a classical article [3] entitled "Can quantum mechanical description of physical reality be considered complete?". In that article, the authors define "elements of physical reality" by the following criterion:*
*If, without in any way disturbing the system, we can predict with certainty ... the value of a physical quantity, then there exists an element of physical reality corresponding to this physical quantity.*
*The criterion is then applied by EPR to a composite quantum system consisting of two distant particles, with an entangled wave function such that a measurement performed on one of the particles allows to predict with certainty the results of a similar measurement that can (but need not) be performed on the other, distant particle. It then follows from an analysis of these conceptual measurements that more information potentially exists than can be supplied by the wave function, and thus EPR " are forced to conclude that the quantum mechanical description of physical reality given by wave functions is not complete".*

Asher Peres continues :

*The simplest example of such a situation is that of two spin ½ particles, far apart from each other, in a singlet state. With the standard notations of Pauli matrices, we have*

$(E_{01} + E_{10})\psi = 0$

$(E_{02} + E_{20})\psi = 0$ (8)

Actually Peres uses the notations $\sigma_{1j}$ and $\sigma_{2j}$ ($j = x, y$) to represent spin operators for particles 1 and 2. We used instead the algebraic notations $E_{0j}$ and $E_{j0}$ ($j = 1,2,3$) that are respectively the algebraic basic elements that we expressed in (6). In this manner $E_{01}, E_{02}, E_{03}$ are the three basic elements relating the spin of the first particle, and $E_{10}, E_{20}, E_{30}$ are those relating the other particle.

Asher Peres continues:

*The first equation in (8) asserts that measurements of $E_{01}$ and $E_{10}$, if performed, shall yield opposite values, $m_{1x}$ and $m_{2x}$, respectively. Each one of these operators thus corresponds to an " element of reality" because its value can be ascertained, without perturbing in any way the particle to which this operator pertains, by means of a measurement performed on the other particle. The same interpretation can be given to the second equation in (8).*

Asher Peres continues:

*…(in) our example of a pair of spin ½ particles in a singlet state, w may define the numerical value of the product $E_{01}E_{20}$ as the product of the individual numerical values $m_{1x}m_{2y}$. Likewise, the numerical value of $E_{10}E_{02}$ is the product $m_{2x}m_{1y}$. From the foregoing discussion, these products must be equal; but, on the other hand, they must be opposite, because the singlet state also satisfies*
$(E_{01}E_{20} + E_{10}E_{02})\psi = 0$ (9)
*What we have here is no longer a paradox, but an algebraic contradiction.*

We conclude: it is an algebraic contradiction that is at the origin of the paradox.
Our attempt is now to give a quantum like explanation and solution of such algebraic contradiction and paradox.

## 4. The Interpretation and the Solution of the EPR Paradox.

We must now return to use our algebraic quantum like rough scheme of quantum mechanics.
One advantage that we know have is that we may now use new algebraic elements to represent combined algebraic elements.
In fact, we may now introduce such new basic elements that result from their previous combination
$E_{11} = E_{01}E_{10} = E_{10}E_{01}$ ; $E_{22} = E_{02}E_{20} = E_{20}E_{02}$ (10)
and
$E_{33} = E_{03}E_{30} = E_{30}E_{03}$ (11)
According to the paradox, we attribute to $E_{11}$, to $E_{22}$ and to $E_{33}$ the numerical value of -1. In our algebraic quantum like scheme this is to say that we have an idempotent $\psi$ such that
$(E_{11} + 1)\psi = 0$ ; $(E_{22} + 1)\psi = 0$ ; $(E_{33} + 1)\psi = 0$ (12)
The idempotent $\psi$ is $\psi_1\psi_2\psi_3$ with
$\psi_1 = \frac{(E_{11} - 1)}{2}$ ; $\psi_2 = \frac{(E_{22} - 1)}{2}$ ; $\psi_3 = \frac{(E_{33} - 1)}{2}$ (13)
and
$E_{11}\psi = -\psi$ ; $E_{22}\psi = -\psi$ ; $E_{33}\psi = -\psi$ (14)
In addition we have that
$\psi_1\psi_2\psi_3 = \psi_2\psi_1\psi_3 = \psi_3\psi_1\psi_2$ (15)
Note that in our quantum like algebraic scheme we operate by idempotents that in some manner may resemble traditional wavefunctions and quantum eigenstates of the orthodox quantum mechanics. However, idempotents are expressed here by the same basic elements of the algebra and thus, in conclusion, we have not the introduction of new quantities or new physical entities operating in the inner of the formulation but all the approach remains always confined in the inner of the algebraic formalism itself.

We must now describe all what happens during the measurements on the singlet state of the two particles in the framework of the EPR. Note that, owing our approach, we must arrive to give an algebraic arrangement about all that happens during an EPR experiment. We must able to characterize and/or to translate all the physical phenomenology in some final algebraic expressions. This is the purpose that we intend to reach.

First of all, through our quantum like scheme, we are now in the condition to evidence from an algebraic point of view the reason by which the vision of reality that the authors of EPR introduced, results so ingénue and strongly linked to the traditional approach based on our direct experience. This is the same reason by which Asher Peres found algebraic contradictions by the (8) and the (9) previously formulated, and, finally, it is the same motivation by which often quantum mechanics results to our eyes so actually distant from the language of our ordinary experience.

Let us look again to the (7). As previously said, they represent all the algebraic sets that we may form when we consider basic elements not at the generic order n=2 but at n=4. Each set of basic elements satisfies the two basic axioms-relations given in (1) and (2) in the sense that the square of each basic element in each set is 1 and each considered triple win of basic elements in each set, satisfies basic non commutation relations as given in (2).

The other important feature is that, when we pass from the most simple basic set that is formed by the pure basic elements as $(E_{01}, E_{02}, E_{03})$ or $(E_{10}, E_{20}, E_{30})$ to more complex sets in the sense that there are now combined basic elements as $E_{ii}$ or $E_{ij}$ (see still the (7) ), profound links appear among such sets: as example, looking at the (7), the basic element $E_{12}$ will appear simultaneously in the first set, in the seventh set, in the thirteenth set, and in the sixteenth set. A similar thing will happen for the other combined $E_{ii}$ or $E_{ij}$ remaining basic elements. In brief, not only each set is regulated by its regime of non commutativity linked to its proper regime of $(i,j,k)$ permutations of (1,2,3) but, in addition, such sets result profoundly linked among them, and this is to say that they realize a very complex regime of functional dependencies among them.

If we look the thing under the profile of a possible quantum like physical interpretation, each algebraic set fixes a regime of potentialities and they results profoundly linked or functionally dependent. To be more rigorous!. Let us consider again the basic elements at order n=2 , that is to say the (1) and the (2). A member of this algebra is

$q = x_0 + x_1 e_1 + x_2 e_2 + x_3 e_3$

with $q \in Cl_3$ and $x_i$ $(i = 0,1,2,3)$ number in some field. In this manner we identify a space, a quantum like space that we could call the abstract space of the potentialities. Going to the (7) we have in a similar way an abstract space of potentialities each time we consider one of the sets given in (7). The previous mentioned functional dependence among the sets configures a final structure of abstract spaces of potentialities that realize of course functional dependence.

The conclusion to which one arrives , seems therefore decidedly expected.

The quantum like picture of physical reality as translated by a simple algebraic formulation results after all more and more complex of the traditional framework that one may delineate thinking only on the basis of our ordinary experience. In particular, considering the case of single or combined measurements, as it was in the case of EPR and, in particular, in the case of the EPR formulation as it was given by Asher Peres and as we reported it in the previous section, in no case it results licit to expect something of similar between single and combined measurements as it was discussed previously by the (8) and the (9) in the previous section.

From an algebraic point of view it is the complex picture of non commutativity and of functional dependence, as previously explained, that every time in a measurement renders the physical context of the measure very specific and very pertinent to the context- measurement itself.

This completes our algebraic approach to the EPR argument.

We may now re-consider the (10-15) and, as previously outlined, we may attempt to give an algebraic characterization of the EPR. In (12-15) only combine basic elements appear. Therefore, our only possibility to characterize EPR in an algebraic manner is to analyze them by the single and combined basic elements that are $E_{i0}, E_{0i}, E_{ii}$, and $E_{ij}$. Starting with the (14) that, as we remember again, represents the core of the EPR argument, we start to calculate:

$$E_{0i}(E_{11}+1)\psi = 0 \; ; \; E_{i0}(E_{11}+1)\psi = 0 \; ; \; E_{ii}(E_{11}+1)\psi = 0 \; ; \; E_{ij}(E_{11}+1)\psi = 0 \quad (16)$$

We will calculate also similar expressions for $E_{22}$ and still for $E_{33}$.
After such calculations we obtain some important results. They are given in the following manner:

$$E_{01}\psi = -E_{10}\psi = -iE_{23}\psi \; ; \quad (17)$$
$$E_{02}\psi = -E_{20}\psi = iE_{13}\psi \; ; \quad (18)$$
$$E_{03}\psi = -E_{30}\psi \; ; \; E_{03}\psi = iE_{21}\psi \; ; \quad (19)$$
$$E_{12}\psi = -E_{21}\psi \; ; \quad (20)$$
$$E_{23}\psi = -E_{32}\psi \; ; \quad (21)$$
$$E_{13}\psi = -E_{31}\psi \; ; \quad (22)$$

Remember that the (17-22) represent only algebraic calculations in a rough quantum like scheme. However, looking step by step to each equation given from the (17) to the (19), one verifies that they synthetize with extraordinary mathematical and physical rigor all that happens during EPR. In addition the (17-22) add also new equations, formally and conceptually unknown in traditional quantum mechanics, and they give the results of such articulated phenomenology of non commutativity and of dependence among the involved algebraic sets as we articulated it previously. Note that all the equations obtained from the (17) to the (22) are direct algebraic consequence of the (14) that delineate our EPR. Therefore, the (17-22) represent all the algebraic quantum like relations that must potentially hold as counter part of a quantum system of two ½ spin particles in a singlet state. Let us observe that in seventy years of research on EPR it was never evidenced that in EPR systems, obeying to the (14), the (17-22) must hold at all.
Let us examine in detail some equations.
The conclusion that
$$E_{0i} = -E_{i0} \text{ (with respect to } \psi\text{)} \quad (23)$$
is in perfect accord with the physical features of the considered EPR quantum system.
We have also that always it must be
$$E_{01} = -iE_{23}; \; E_{02} = iE_{13}; \; E_{03} = iE_{21} \quad (24)$$
We have also that
$$E_{13} = -E_{31} \quad \text{and} \quad E_{23} = -E_{32} \quad (25)$$
Finally, we have also the most relevant equation for our work. It must be that
$$E_{12} = -E_{21} \quad (26)$$
Its particular relevance does not arise from particular notations of algebraic or physical importance but from the fact that Asher Peres, first trough the (8) and thus by the (9) arrived to formulate the EPR paradox or the algebraic contradiction as he called it correctly. We have to explain the (26) if we aim to give solution to the paradox. It should be now clear on the basis of the arguments that we discussed previously, but here we intend to give a further confirmation and elucidation of the correctness of our arguments.
Let us consider the basic elements $E_{12}$ and $E E_{21}$. From the algebraic context that we have introduced until here, it is evident that we may express such two basic elements in different manner.

Among such different representations, the most simple, clear, but also responding to its counter part to a too ingénue realistic vision, is to write such two elements in their basic forms that are:

$$E_{12} = E_{10}E_{02} \quad \text{and} \quad E_{21} = E_{20}E_{01} \tag{27}$$

In this case we are ingenuously admitting as existing " elements of reality" but we are ignoring the complexity of such reality as previously explained through the articulated picture of non commutativity and functional dependence of sets that instead may characterize our system. And in fact we observe that immediately the (27) goes in contradiction with itself since, observing the (23), we have as example that $E_{01} = -E_{10}$ and $E_{02} = -E_{20}$
and consequently

$$E_{12} = E_{10}E_{02} = E_{01}E_{20} = E_{21} \tag{28}$$

and
$$E_{21} = E_{20}E_{01} = E_{02}E_{10} = E_{12}$$

That is in contradiction with the (26) that instead is rigorously required in our algebraic formulation and in EPR. So we arrive to the conclusion and thus to the explanation of the paradox via our algebraic quantum like formulation.

We are engaged in an algebraic calculation that peremptorily requires to respect its basic features of non commutativity and link among sets. In other terms, in order to obtain that

$$E_{12} = -E_{21} \tag{29}$$

we cannot ignore that they must necessarily contain almost an explicit permutation of (1,2,3). This is to say that peremptorily we must account for non commutativity.

Explicting permutations, this time we obtain

$$E_{12} = \frac{E_{13}E_{01}}{i} \tag{30}$$

Performing the same operation we also obtain that

$$E_{21} = \frac{E_{22}E_{03}}{i} \tag{31}$$

In this manner, $E_{12}$ may be written in the following terms

$$E_{12} = \frac{E_{10}E_{03}E_{01}}{i} = E_{10}\frac{E_{03}E_{01}}{i} \quad \text{and} \quad E_{02} = \frac{E_{03}E_{01}}{i} \tag{32}$$

while $E_{21}$ will be written in the following manner

$$E_{21} = \frac{E_{20}E_{02}E_{03}}{i} = E_{20}\frac{E_{02}E_{03}}{i} \quad \text{and} \quad E_{01} = \frac{E_{02}E_{03}}{i} \tag{33}$$

We may now solve the (32) and the (33). We find that

$$E_{12} = \frac{E_{10}E_{03}E_{01}}{i} = \frac{E_{03}E_{10}E_{01}}{i} = -\frac{E_{03}}{i} \tag{34}$$

and

$$E_{21} = \frac{E_{22}E_{03}}{i} = \frac{E_{20}E_{02}E_{03}}{i} = \frac{-E_{03}E_{02}E_{20}}{i} = \frac{E_{03}}{i} \tag{35}$$

that is to say that

$$E_{12} = -E_{21} \tag{36}$$

and the EPR paradox is solved in its quantum like algebraic version.

## 5. Conclusion

We would conclude outlining the basic foundations of our formulation.
The strong difference between the (28) and the (30) and the (31) resides in the two different models of reality that they show off and that have their essence in the algebraic counterpart of the quantum like algebraic model that we have introduced. The (30) and (31) involve a permutation that instead is missing in (28). In our formulation a permutation is connected with an operation of non commutativity. Therefore, when we omit to consider a permutation, as in the (28), we perform a great error of physical analysis: we consider a commutation that in the corresponding language of traditional quantum mechanics means simultaneous measurability of observables with precision, where instead a non commutation is required and in this case it implies in the standard language of the theory that two observables cannot be simultaneously measured with precision. Thus, in conclusion, in both cases, respectively we design two completely different models of reality. As consequence, a paradox is induced.

*We may conclude outlining the particular importance to have expressed $E_{02}$ and $E_{01}$ as in (32) and in (33).*
*The expression*

$$E_{02} = \frac{E_{03}E_{01}}{i} \quad (37)$$

*in $E_{12}$ given in (28) outlines that, if we consider $E_{10}$ and $E_{12}$ we cannot escape to consider the functional dependence (given by permutation and thus non commutativity) of $E_{02}$ from $E_{01}$ and $E_{03}$.*
*In the same manner, the presence of*

$$E_{01} = \frac{E_{02}E_{03}}{i} \quad (38)$$

*in $E_{21} = E_{20}E_{01}$*
*states that considering $E_{20}$ and $E_{21}$, we cannot escape to consider the functional dependence of $E_{01}$ from $E_{02}$ and $E_{03}$.*

In brief, in this paper mathematics has helped us to conquer the physical explanation of the paradox that we have just indicated. The consequence is that, owing to the regime of functional dependences, as explained in physical terms by the sentences included from the (37) to the (38), Nature does not respond to the ingenuous vision of realism that was admitted in the past. Actually, Nature is more complex than we could have expected.

**References**
[1] E. Conte, GP. Pierri, A. Federici, L. Mendolicchio, J.P. Zbilut, A model of biological neuron with therminal chaos and quantum like features, Chaos, Solitons and Fractals, 30, 774-780, 2006;
E. Conte, A. Khrennikov, J.P. Zbilut, The transition from ontic potentiality to actualization of states in quantum mechanical approach to reality: a proof of a mathematical theorem supporting it, Arxiv.org quant-ph/0607196;
E. Conte, GP. Pierri, L. Mendolicchio, A. Khrennikov, J.P Zbilut, On some detailed examples of quantum like structures containing quantum potential states in the sphere of biological dynamics, Arxiv.org physics/0608236;
E. Conte, O. Todarello, A. Federici, J.P. Zbilut, F. Vitiello, M. Lopane, A. Khrennikov, J.P Zbilut, Found experimental evidence of quantum like behaviour of cognitive entities, Chaos, Solitons and Fractals, 31, 1076-1088, 2007; Arxiv.org quant-ph/0307201; Some remarks on an experiment suggesting quantum like behaviour of cognitive entities and formulation of an abstract quantum mechanical formalism to describe cognitive entity and its dynamics , Arxiv.org physics/0710.592

E. Conte, O. Todarello, A. Federici, F. Vitello, M. Lopane, A. Khrennikov, A preliminar evidence of quantum like behaviour in measurements of mental states, in Quantum Theory: Reconsideration of Foundations 2, 679-702, Vaxjo University Press, 2003;

E. Conte, Quant. Mech., PIT. Press, Italy, 2000.

[2]E. Conte, The formulation of EPR paradox in quantum mechanics, Fundamentals Problems in natural sciences and engineering, Proceedings, Saint-Petersburg, 2002

[3] A. Einstein, B. Podolsky, N. Rosen, Can quantum mechanical description of physical reality be considered complete?, Phys. Rev. 47, 777-781, 1935;

[4] A. Peres, Recursive definition for elements of reality, Found. of Phys., 22, 3, 357-361, 1992.